\documentclass{mn2e}

\usepackage{epsfig}
\usepackage{rotating}
\newif\ifAMStwofonts
\AMStwofontstrue

\title{Spitzer and ISO Galaxy Counts in the Mid-Infrared }
\author[C.P. Pearson]
       {Chris Pearson$^1$\thanks{E-mail: Chris Pearson (cpp@ir.isas.jaxa.jp)  } \\
 $^1$Institute for Space and Astronautical Sciences, Japanese Aerospace eXploration Agency, Sagamihara, Kanagawa 229 8510, Japan}

\date{Accepted .\\
      Received ;\\
      in original form 2004 October }
\pagerange{\pageref{firstpage}--\pageref{lastpage}}

\begin{document}

\label{firstpage}

\maketitle

\begin{abstract}
Galaxy source counts that simultaneously fit the deep mid-infrared surveys at 24 microns and 15 microns made by the Spitzer Space Telescope and the Infrared Space Observatory (ISO) respectively are presented for two phenomenological models. The models are based on starburst and luminous infrared galaxy dominated populations.  Both models produce excellent fits to the counts in both wavebands and provide an explanation for the high redshift population seen in the longer Spitzer  24 micron band supporting the hypothesis that they are luminous-ultraluminous infrared galaxies at z=2-3, being the mid-infrared counterparts to the sub-mm galaxy population. The source counts are characterized by strong evolution to redshift unity, followed by less drastic evolution to higher redshift.  The number-redshift distributions in both wavebands are well explained by the effect of the many mid-infrared features passing through the observation windows. The sharp upturn at around a milliJansky in the 15 micron counts in particular depends critically on the distribution of mid-infrared features around 12 microns, in the assumed spectral energy distribution.
\end{abstract}


\section{Introduction}\label{sec:introduction}

The detection of the peak in the far-infrared background at 140$\umu$m by the COBE-DIRBE  instrument has emphasized that much of the star formation in the Universe has taken place in dust enshrouded environments  and that the galaxies responsible have undergone strong evolution to the present epoch (e.g. Hauser et al. ~\shortcite{hauser98},  Lagache \& Puget~\shortcite{lagache00}). This evolution has also been inferred and confirmed at wavelengths from 7-170$\umu$m by a progression of increasingly inspiring space missions from  {\it IRAS}, {\it ISO} to {\it Spitzer} (Soifer et al. ~\shortcite{soifer87}, Kessler et al.~\shortcite{kessler96}, Werner et al.~\shortcite{werner04} ) and at sub-mm wavelengths by the SCUBA instrument ~\cite{holland99}.  The emission in the dust spectrum of galaxies due to star formation peaks in the wavelength region between 60-100$\umu$m however a significant fraction ($\sim$40$\%$) of a typical star-forming galaxy's luminosity is radiated at mid-infrared wavelengths from 7-40$\umu$m ~\cite{soifer87}. Moreover, unlike the far-infrared, observations at mid-infrared wavelengths are not so severely constrained by detector array size, resolution and detector effects such as transients and glitches. The mid-infrared spectra of star-forming galaxies are dominated by a complex set of emission features at 3.3, 6.2, 7.7, 8.6, 11.3 \& 12.7$\umu$m attributed to Polycyclic Aromatic Hydrocarbon (PAH) molecules and the silicate absorption feature at 9.7$\umu$m (Boulanger et al. ~\shortcite{boulanger98}, Peeters et al. ~\shortcite{peeters04}). These PAH features can contribute 10-30$\%$ of the bolometric infrared luminosity of infra-red galaxies and are good indicators of the star-formation rate ~\cite{spoon04a}. Typical widths of these features (up to10$\umu$m) can be comparable to instrument filter bandwidths  (e.g., 3.5$\umu$m \& 6$\umu$m for the ISOCAM-LW2(5-8.5$\umu$m) filter and ISOCAM-LW3(12-18$\umu$m) filters respectively) and will therefore have a profound effect on any observations in the mid-infrared.

The strongest constraint on the evolution in the mid-infrared until recently, has been the source counts at 15$\umu$m in the ISOCAM-LW3 band. The {\it ISO} counts span over 4 orders of magnitude in flux ~\cite{elbaz99}. The flux density range from 150-0.5mJy is covered by the European Large Area {\it ISO} Survey over four main fields over a total of $\sim$12sq.deg. (ELAIS, Oliver et al. ~\shortcite{oliver00}). At flux densities $>$2mJy the source counts are Euclidean in nature and are well fitted by a combination of star-forming and quiescent galaxies with little or no contribution from more extreme luminous (LIG, $L_{IR}>10^{11}L_{\sun} $) or ultraluminous (ULIG, $L_{IR}>10^{12}L_{\sun} $) infrared galaxies (Serjeant et al. ~\shortcite{serjeant00}, Gruppioni et al. ~\shortcite{gruppioni02} Rowan-Robinson et al. ~\shortcite{mrr04}). At fluxes $<$2mJy the source counts show a strong departure from non-evolving scenarios. The fainter {\it ISO} surveys detected extreme evolution producing super Euclidean slopes at fluxes $< 1mJy$, flattening at fainter fluxes of $0.4mJy$ (Fadda et al. ~\shortcite{fadda04}, Pozzi et al.~\shortcite{pozzi04}, Metcalfe et al.~\shortcite{metcalfe03}, Gruppioni et al. ~\shortcite{gruppioni02},  Altieri et al.~\shortcite{altieri99}, Aussel et al..~\shortcite{aussel99}) The majority of the deep ISOCAM sources have been identified with luminous infra-red galaxies with luminosities of the order of $10^{11}L_{\sun} $ (Flores et al. ~\shortcite{flores99}). The {\it ISO} 15$\umu$m counts are well fit by models assuming a rapid increase in the star formation activity between 0$<$z$<$1 (e.g.  Pearson~\shortcite{cpp01a}, Xu et al. ~\shortcite{xu03}, Lagache et al.  ~\shortcite{lagache03}). The MIPS 24$\umu$m band on the {\it Spitzer} Space Telescope is well suited to the study of distant star forming galaxies at z$>$1 since the 24$\umu$m band is sampling the rest frame PAH emission from 6-12$\umu$m from 0.7$<$z$<$2.5. Observations in the 24$\umu$m band over the flux range 50-0.06mJy have confirmed the strong evolution of the mid-infrared population (Papovich et al. ~\shortcite{papovich04}, Marleau et al. ~\shortcite{marleau04}), but however cannot be explained by evolutionary models based solely on the previous {\it ISO} 15$\umu$m population (e.g. Xu et al. ~\shortcite{xu03}, Chary \& Elbaz ~\shortcite{chary03}, King \& Rowan-Robinson ~\shortcite{king03}, Lagache et al.  ~\shortcite{lagache03}). The {\it ISO} 15$\umu$m  differential counts peak at around 0.4mJy with a median redshift $\sim$0.8 (Elbaz et al. ~\shortcite{elbaz99}, Elbaz et al. ~\shortcite{elbaz02}), conversely the peak in the 24$\umu$m counts occurs at flux densities almost twice as faint around 0.2mJy implying a higher redshift distribution for the dominant sources ~\cite{chary04} for a realistic range of the 24$\umu$m/15$\umu$m flux ratio ~\cite{gruppioni05}. 

In this letter the emphasis is on the simultaneous fitting of the source counts in the  {\it ISO}-ISOCAM-LW3 15$\umu$m and {\it Spitzer}-MIPS 24$\umu$m bands. Detailed results on the modelling of the spectral energy distributions, galaxy colours, discussion of the galaxy populations and source counts in the shorter {\it Spitzer}-IRAC bands and the longer wavelength MIPS bands (70$\umu$m \& 160$\umu$m)  will be given in a later contribution ~\cite{cpp05}. In section~\ref{sec:model}  two phenomenological galaxy evolution models used to model the midinfrared counts are presented. In section~\ref{sec:counts} we discuss the resulting fits to the source counts at 15$\umu$m and 24$\umu$m. The conclusions are given in section~\ref{sec:discussion} . A concordance cosmology of  $H_o = 72kms^{-1}Mpc^{-1}, \Omega=0.3, \Lambda=0.7$ is assumed.

\section{The mid-infrared population}\label{sec:model}

The models presented are referred to as the {\it Bright End} evolution model and the {\it Burst} evolution model and are modified and updated realizations of those presented in Pearson ~\shortcite{cpp01a}, Pearson \& Rowan-Robinson ~\shortcite{cpp96}.
The previous models were based on the far-infrared 60$\umu$m luminosity function of PSCz  {\it IRAS} galaxies ~\cite{saunders00}. The new models use the 15$\umu$m type dependent luminosity  functions derived from  the {\it ISO}-ELAIS survey of Pozzi et al. ~\shortcite{pozzi04} to represent the normal + starburst/LIG/ULIG populations and Matute et al.~\shortcite{matute02}   to represent the Type-1 AGN (including Seyfert 1). Since Matute et al.~\shortcite{matute02} only provide a parametric luminosity function for the type 1 population, the 12$\umu$m luminosity function of Rush et al.~\shortcite{rush93} is retained for the type 2 population. The use of mid-infrared luminosity functions as opposed to the more traditional far-infrared variants to model the source counts at 15 \& 24$\umu$m minimize any errors due to assumed SED shape and K-corrections. The luminosity function of Pozzi et al. ~\shortcite{pozzi04}  is used in preference to the 12$\umu$m \& 25$\umu$m  {\it IRAS} luminosity functions of Xu et al.~\shortcite{xu01} or Fan et al. ~\shortcite{fang98} because the former is segregated specifically into normal quiescent and starburst populations based on {\it ISO} colour criteria .

The adopted SED is the most important factor in the models. The complicated nature of the mid-infrared spectra and the variation between models, even for more normal galaxies can have a dramatic effect on the derived source counts of a galaxy population. In the final analysis of the ELAIS survey, Rowan-Robinson et al. ~\shortcite{mrr04} have shown that the infrared galaxy population can be divided into 4 distinct spectral classes of normal quiescent (cirrus) galaxies, starbust (M82) galaxies, LIG/ULIG ($lgL_{IR} > 10^{11-12}L_{\sun} $) sources (ARP220) and AGN (dust torus).  

The original normal galaxy used in Pearson ~\shortcite{cpp01a}  has been replaced by the SED of Dale et al. ~\shortcite{dale01} which incorporates a more recent dust model (see Pearson \& Efstathiou ~\shortcite{cppef05} for a comaprison of cirrus galaxy SEDs). Although, for the normal galaxies, the effect of changing the SED in the 24$\umu$m band is minimal, at 15$\umu$m the original model produced excess power in the differential counts at the level of a few mJy due to extremely strong and narrow PAH emission, masking the evolutionary upturn in the differential counts at that level. 

For the starburst template, the critical point in the SED is where the Wien-side of the dust emmission curve meets the mid-infrared PAH emmission. Analysis of model starburst SEDs shows a large variation in resulting K-corrections due to 2 factors, the slope of the SED at this point (an increasing slope to shorter wavelengths producing a favourable negative K-correction) and the inclusion of distinct 12.7$\umu$m \& 11.3$\umu$m features to produce a more accurate distribution of mid-infrared features  (Boulanger et al.~\shortcite{boulanger98}, Diedenhoven et al.~\shortcite{diedenhoven04}). Thus we have retained the original Starburst (M82 like) SED of Efstathiou, Rowan-Robinson \& Siebenmorgen ~\shortcite{esf00} but have replaced the SED in the range from 18-5$\umu$m, with the observed ISOCAM Circular Variable Filter spectrum of Forster Schreiber et al. ~\shortcite{forster03}. Note that a similar adjustment was necessary in the models of Lagache et al.  ~\shortcite{lagache04} in order to produce a reasonable fit to the {\it Spitzer}  24$\umu$m counts.

 The luminous infrared galaxy population is represented by the  archetypical ARP220 SED of Efstathiou, Rowan-Robinson \& Siebenmorgen ~\shortcite{esf00} which gives a good fit of the spectrum of ARP220 from sub-mm to optical wavelengths and is in reasonable agreement with the results of Spoon et al. ~\shortcite{spoon04b}. Note that although it is known that ARP220 has an atypical far-infrared/mid-infrared luminosity ratio, we justify its use by the fact that it provides good fits to many of the ELAIS sources ~\cite{mrr04} and has more typical 6.7-12-15$\umu$m luminosity ratios ~\cite{elbaz02}. 
 
 The AGN template uses the dust torus model of Rowan-Robinson ~\shortcite{mrr95} extended to near-infrared-optical wavelengths by King \& Rowan-Robinson ~\shortcite{king03}. All the assumed SEDs are found to be consistent with the {\it ISO}-ELAIS colours of Rowan-Robinson et al. ~\shortcite{mrr04}. For the calculation of the source counts the SEDs are smoothed by the respective filter bandwidths corresponding to the {\it ISO}-LW3 15$\umu$m and {\it Spitzer}-MIPS 24$\umu$m bands.

 We investigate two particular models,  categorized by their dominant populations of  starburst (M82 like)  and ULIG (Arp 220 like) sources respectively.   The  {\it Bright End} Model broadly follows the evolutionary  scenario of Pearson \& Rowan-Robinson ~\shortcite{cpp96} assuming the starburst/AGN  and LIG/ULIG population luminosity function $\Phi (L,z)$ evolves in both luminosity, $k$, and density, $g$, as $\Phi (L/(1+z)^{k},0).(1+z)^{g}$, with $k=3.3$ \& $2.5$ and $g=3.3$ \& $3.5$ for the starburst and LIG/ULIG components respectively to z$\sim$1 and constant thereafter. The {\it Burst} Model broadly follows the evolutionary scenario  of Pearson  ~\shortcite{cpp01a} and assumes a similar power law  evolution in luminosity and density, for the starburst and AGN populations and an  initial exponential burst + power law evolutionary scenario  for the  ULIG population. In this scenario, the starburst population evolves as $(1+z)^{k,g}$, $k=g=3.2$ while the more luminous galaxies evolve exponentially to z$\sim$1 in density equivalent to  $g\sim 7$ and then as a lower power law in both density and luminosity to higher redshifts. This violent evolutionary phase from z$\sim$0-1 is consistent with extrapolations from the  {\it IRAS} faint source survey (Bertin et al.~\shortcite{bertin97}, Kim \& Saunders ~\shortcite{kim98}), more extreme models of evolution in the {\it ISO}  population ~\cite{lagache03}, models of bright sub-mm SCUBA  galaxies (Smail et al.~\shortcite{smail97}, Hughes et al.~\shortcite{hugh98}, Scott et al. ~\shortcite{scott02}, Mortier et al. ~\shortcite{mortier04}) and the most recent models of the {\it Spitzer} far-infrared poulation ~\cite{lagache04}. In both models the normal galaxies are assumed to be non-evolving.

\section{Counts and Redshift Distributions}\label{sec:counts}

\begin{figure}
\centering
\centerline{
\psfig{figure=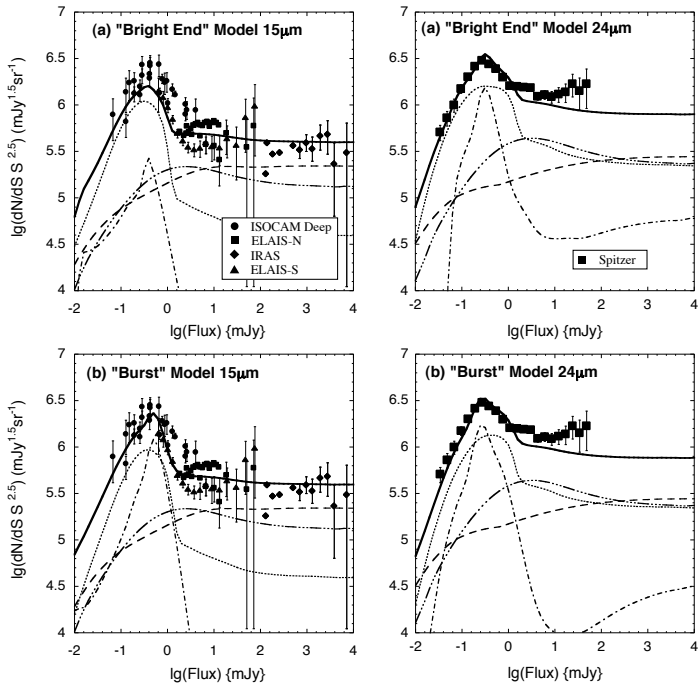,height=9cm}
}
\caption{Model fits to source counts in the ISOCAM 15$\umu$m and {\it Spitzer} 24$\umu$m bands for the {\it Bright End} (a) \& {\it Burst} (b) models. Counts are presented in differential form, normalized to the non-evolving Euclidian universe. Total counts are shown with individual components corresponding to Normal (dash), Starburst (dot), LIG/ULIG (dot-dash) sources and AGN (triple dot-dash). {\it ISO} 15$\umu$m source counts from faint to bright fluxes are from  the {\it ISO} cluster lens survey - Altieri et al. (1999), Hubble Deep Field - Oliver et al. (1997), Aussel et al. (1999)), Lockman Hole - Elbaz et al. (1999), ELAIS-S - Gruppioni et al (2002), ELAIS-N - Serjeant et al. (2000) and normalized  {\it IRAS} counts from 12$\umu$m - Rush et al. 1993. The {\it Spitzer}-MIPS  24$\umu$m source counts are from Papovich et al. (2004).
\label{counts}}
\end{figure}

In figure~\ref{counts} the model fits to the observed counts in  the {\it ISO} 15$\umu$m and {\it Spitzer} 24$\umu$m wavebands are shown for the {\it Bright End} (a)  \&  {\it Burst}  (b) evolutionary scenarios respectively. The counts are shown in Euclidean normalized differential form. The two models are capable of simultaneously fitting the differential source counts in both the {\it ISO} 15$\umu$m and {\it Spitzer} 24$\umu$m wavebands. The fits at 15$\umu$m confirm the strong evolution to z$\sim$1 implied by the {\it ISO} surveys while the counts at fainter fluxes at 24$\umu$m indicate a population of sources at higher redshift. The more conventional  {\it Bright End} model predicts that star-forming galaxies rather than ULIG sources dominate  the  15$\umu$m source counts.  

Pozzi et al. ~\shortcite{pozzi04} have fitted the  ELAIS-S 15$\umu$m counts assuming an evloving starburst population with $(1+z)^{k,g}$, $k=3.5, g=3.8$. Although assuming the Pozzi et al. ~\shortcite{pozzi04} evolution for the starburst component presented in this work does indeed provide a good fit at 15$\umu$m, it also produces an excess at around 1mJy in the  24$\umu$m band, thus counts at fluxes of around 0.5-2mJy, in the {\it Spitzer}  24$\umu$m band can provide good constraints on the z$<$1 evolution. 

In contrast, the more extreme burst model predicts that the upturn at mJy level fluxes at  15$\umu$m and the peak of the counts at  24$\umu$m are attributed to the emergence of a new population of LIG/ULIG sources. Finally, both the {\it Bright End} Model and the {\it Burst} Model predict that the peak in the 24$\umu$m differential counts at $\sim$0.2mJy is produced by the luminous LIG-ULIG population consistent with the early analysis of the {\it Spitzer}  24$\umu$m  data (Yan et al. ~\shortcite{yan04}, Le Floc'h et al. ~\shortcite{lefloch04}) . 

Thus, although from figure~\ref{counts} both models are in agreement as to the origin of the dominant source population at 24$\umu$m, they differ in their prediction for the source of the upturn (note, not the peak) in the differential counts at  15$\umu$m, thus it is the counts at  15$\umu$m that have a better chance of distinguishing between the two evolutionary scenarios. For the  {\it Burst} model, we would expect to see a significantly higher fractional contribution of LIG/ULIG sources (an order of magnitude in the differential counts at the mJy level) that simply are not appreciably present in the {\it Bright End} model. 

Note that at the depth of the ELAIS survey ($S_{15}>0.7mJy$) the predictions of both models are consistent with the near-infrared and optical imaging of the ELAIS-S field by Pozzi et al.  ~\shortcite{pozzi03}, with the majority of 15$\umu$m sources being luminous starburst galaxies and approximately 20$\%$ of sources attributed to AGN (Type 1 \& 2). Rowan-Robinson et al. ~\shortcite{mrr04} find a high proportion $\sim$14$\%$ of ULIGs in the ELIAS final band merged catalogue. This proportion is $\sim$10$\%$ for the {\it Bright End} model. The  {\it Burst} model predicts a fraction of $\sim$30$\%$ which is still consistent with the ELAIS results if the blank optical identifications ($\sim$10$\%$) can be attributed to luminous infrared galaxies. Spectroscopy of  the ELAIS-S sources has been made by La Franca et al. ~\shortcite{lafranca04}. Of $\sim$87$\%$  of sources spectroscopically identified, two populations dominated the 15$\umu$m sources over the range 0.5-150mJy, with approximately three quarters contribution coming from normal to extreme starbursts and the remaining contribution coming from AGN. Around $\sim$18$\%$ of the ELAIS-S sources have no counterpart to optical maginitudes R$>$23. Moreover  La Franca et al. ~\shortcite{lafranca04} have shown that it is these sources are responsible for the upturn in the differential sources counts around $\sim$1mJy at 15$\umu$m and that they are expected to have luminosities in the $10^{11}-10^{12}L_{\sun} $ range, i.e. luminous and ultra-luminous infra-red galaxies.

\begin{figure}
\centering
\centerline{
\psfig{figure=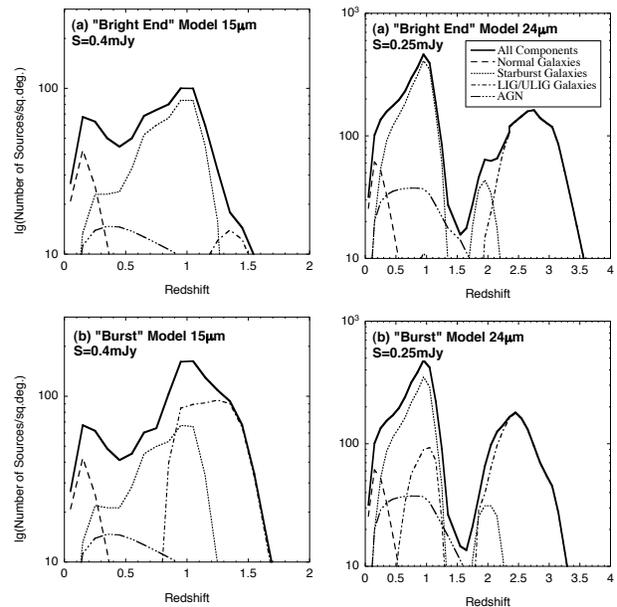,height=8cm}
}
\caption{Model Number-Redshift distributions for ISOCAM 15$\umu$m and {\it Spitzer} 24$\umu$m bands  for the {\it Bright End} (a) \& {\it Burst} (b) models. The distribution are per sq.deg. with redshift bin size $\delta z=0.1$.  Total numbers are shown with individual components corresponding to Normal, Starburst, LIG/ULIG sources and AGN. The distributions are for $S_{15}=0.4mJy$ \& $S_{24}=0.25mJy$ and   correspond to the approximate peaks in the differential counts given in figure 1.
\label{nz}}
\end{figure}  

From figure ~\ref{counts}, it can be seen that although both the fainter 15$\umu$m \& 24$\umu$m differential counts can be well fitted at the fainter fluxes, the brighter counts are slightly underestimated in the  {\it Spitzer} 24$\umu$m band. This could be attributed to some variance in the  {\it Spitzer} fields since the initial counts from the SWIRE survey are somewhat lower at fluxes brighter than 10mJy (Shupe et al., in preparation). Alternatively, this effect could be due to a low normalization in the assumed ELAIS-S field luminosity function since although there is a good agreement with the evolution found in the deeper {\it ISO} surveys, including the upturn in the differential counts, in the brighter flux range of 0.5-10mJy, the counts in the ELAIS-S field are systematically lower than the (corrected) Northern ELAIS fields and the fainter ISO surveys  (Serjeant et al. ~\shortcite{serjeant00}, Vaisanen et al. ~\shortcite{vaisanen02}, Vaccari et al. ~\shortcite{vaccari04}, Rodighiero et al. ~\shortcite{rodighiero04}). 

Both evolutionary models are found to be broadly consistent with the 15$\umu$m redshift distribution at bright fluxes with a median redshift consistent with that found for the ELAIS sources (Pozzi et al. ~\shortcite{pozzi04}, Rowan-Robinson et al.  ~\shortcite{mrr04}). However, we predict an additional component at higher redshift $\sim$1 in line with the assumptions of La Franca et al. ~\shortcite{lafranca04} but not present in the estimations of  Rowan-Robinson et al.  ~\shortcite{mrr04}.  In figure ~\ref{nz} the number-redshift distributions at fainter fluxes corresponding to the approximate peaks in the differential counts at $S_{15}=0.4mJy$ \& $S_{24}=0.25mJy$ are shown respectively for the ISOCAM 15$\umu$m \& {\it Spitzer} 24$\umu$m bands  for the {\it Bright End} (panel a) \& {\it Burst} (panel b) evolutionary models. At the flux corresponding to the peak in the 15$\umu$m differential counts we find good agreement with the fainter {\it ISO} surveys that predict a median redshift between z$\sim$0.8-1.0 and domination by populations of luminous infrared sources (Elbaz et al.~\shortcite{elbaz02}, Oliver et al. ~\shortcite{oliver02}, Mann et al.~\shortcite{mann97}.  At this redshift (z$\sim$0.95) the 7.7$\umu$m rest frame feature passes through the {\it ISO} 15$\umu$m  band producing strong K-corrections, which combined with the strong evolution produces the redshift peak. The effect of the 9.7$\umu$m silicate absorption feature is also prominent at z$\sim$0.5 and is consistent with the drop in density of galaxies observed in the ELAIS-S field at this redshift  ~\cite{lafranca04}. 
At 24$\umu$m a bimodal redshift distribution is seen at z$\sim$1 \& $\sim$2.5 corresponding to the passage of the 6.2-7.7$\umu$m and 11.3-12.7$\umu$m features respectively. A similar bimodal distribution has also been suggested by Chary et al. ~\shortcite{chary04} although the models presented here predict a greater contribution from  higher luminosity infrared sources. A significant  deficit of sources is seen at z$\sim$1.5 which can be attributed to the effect of the silicate feature at 9.7$\umu$m. Although the two evolutionary scenarios predict a different origin for the dominant population at low redshifts, both models are unanimous in predicting a new population of infra-red luminous sources in the redshift range,  z=1-3. The SCUBA sub-mm population would seem likely candidates for this population of sources, indeed observations in the MIPS 24$\umu$m band of SCUBA sources in the Lockman Hole detected almost all of the sources in the range z$\sim$0.5-3.5 ~\cite{egami04}. Furthermore, the redshift distribution of the high-z {\it Spitzer} sources and their predicted luminosity range (LIG-ULIG) correlates well with the median redshift of 2.2-2.5 found for sub-mm sources (Chapman et al. ~\shortcite{chapman03}, ~\shortcite{chapman04}, Frayer et al.  ~\shortcite{frayer04}). Luminous/ultraluminous galaxies are responsible for the peak at around 0.25mJy in the 24$\umu$m differential counts in both models. At this flux level both models predict of the order of more than 1000 such sources per square degree and typical sub-mm fluxes at 850$\umu$m of 3-5mJy over the redshift range 1-2.5, in very good agreement with the observed submillimetre source density in this flux range observed by SCUBA ~\cite {blain99}.

\section{Conclusions}\label{sec:discussion}

The two evolutionary models presented are capable of simultaneously fitting both the 15$\umu$m  {\it ISO} and 24$\umu$m {\it Spitzer} normalized differential source counts. Careful consideration has been given to the assumed distribution of PAH features. The {\it ISO} counts are well fitted by starburst or LIG-ULIG sources strongly evolving to redshift $\sim$1, depending on the model, while both models are unanimous in predicting a new population of luminous infrared galaxies at z$\sim$2-3 detected by the {\it Spitzer} observations at 24$\umu$m, that were previously unexpected from fits to the {\it ISO} counts. These higher redshift sources are interpreted as being the infrared counterparts to the luminous sub-mm population. The number-redshift distributions at 15$\umu$m \& 24$\umu$m, while confirming the strong evolution in the infrared population, show a bimodal distribution dominated by PAH emission features between 3-13$\umu$m and the silcate absorption feature at 9.7$\umu$m. 

Although the original model of Pearson ~\shortcite{cpp01a} provided an excellent fit to the source counts at 15$\umu$m and other ISO wavelengths, like other pre-{\it Spitzer} models it fails to reproduce the bump in the  {\it Spitzer} 24$\umu$m counts at fainter fluxes (see Papovich et al ~\shortcite{papovich04} for a comparision of pre- {\it Spitzer} models). In summary, the main difference in the assumed evolution in the new model formulation and that of the original models is an increase in the evolutionary strength in the 1$<$z$<$2 region. Alternative post-{\it Spitzer} evolutionary models of Gruppioni et al. ~\shortcite{gruppioni05} and  Lagache et al.  ~\shortcite{lagache04}  can be broadly categorized by the  {\it Bright End} evolution model and the {\it Burst} evolution model respectively. All of these models have required some adjustment of the mid-infrared Starburst spectral energy distribution in order to fit the observations. Both the models of Lagache et al. and the  {\it Burst} evolution model are characterized by extreme evolution in the high luminosity infrared galaxy population (i.e. the LIG-ULIGs). These models predict that the dominant population at 15$\umu$m will be starburst galaxies while at 24$\umu$m they will be sources with infrared luminosities of the order of a few $10^{11}-10^{12}L_{\sun} $ at redshifts 1$<$z$<$3. In contrast the model of Gruppioni et al. ~\shortcite{gruppioni05} and similarly the {\it Bright End} evolution model presented here, propose more traditional power law evolution and that the more numerous population at both 15 and 24$\umu$m will be the starburst galaxies, in fact the model of Gruppioni et al. ~\shortcite{gruppioni05} assumes that the contribution of more extreme IR galaxies (ULIGs) at both 15 and 24$\umu$m is negligible.

Deeper observations over larger areas in the 15$\umu$m band and photometric redshifts using mid-infrared multiband observations of PAH emission and the silicate absorption features can break the degeneracy in the models by revealing the dominant population at fainter 15$\umu$m fluxes (Starburst, LIGs, ULIGs). {\it Spitzer} has the ability for imaging in this wavelength range  (16 \& 22$\umu$m) by using the peak-up camera of the infrared spectrograph (IRS, Houck et al. ~\shortcite{houck04}), and Charmandaris et al. ~\shortcite{charmandaris04} have discussed the potential for estimating redshifts and selecting galaxies at z$>$1.5 from the mid-infrared features by using the IRS. Takagi \& Pearson ~\shortcite{takagi05} have shown that a combination of mid-infrared filters can detect luminous infrared galaxies at z$\sim$1-2 due to the strong silicate absorption expected in these sources ~\cite{armus04}. 

The  {\it Spitzer} peak up imager is however, limited by its small field of view (0.9\arcmin x1.3\arcmin). Deep, larger area observations in this wavelength range will be available from the Japanese infrared mission {\it ASTRO-F}  due for launch in late 2005  (Murakami ~\shortcite{murakami04}, Pearson et al. ~\shortcite{cpp04}).  The {\it ASTRO-F} focal plane instruments, include an infrared camera (IRC) with a field of view of $>$10\arcmin x 10\arcmin (almost 100 times the area of {\it Spitzer}'s peak up imager, allowing coverage of the deepest  {\it Spitzer} fields in a single pointing), covering the near-mid-infrared in 9 bands from 2-24$\umu$m and covering the {\it Spitzer wavelength desert} in 4 extra bands  at 9, 11, 15, 20$\umu$m, making it the ideal complement to  {\it Spitzer}.  Deep surveys are planned over sub-few square degree scales with sensitivities down to   0.04mJy at 15$\umu$m ~\cite{cpp01b}, i.e. deep enough to directly probe significant samples of the sources responsible for the upturn in the 15$\umu$m counts. Takagi \& Pearson~\shortcite{takagi05} predict of the order of 1000 candidate LIG/ULIGs could be detected per square degree using the  {\it silicate-break} method which will be a large enough sample to break the degeneracy in the models via specific colour cuts utilizing ASTRO-F's larger number of filters in the critical wavelength range that samples the PAH emission in galaxies from 0$<$z$<$2.5 which is will be only partially possible and  difficult \& time consuming to achieve with  {\it Spitzer}'s peak up imager.  

In the next few years, multiwaveband observations with {\it Spitzer} and {\it ASTRO-F} probing different redshift regimes will examine the infrared active epoch z$\sim$0.5-1.5 (Franceschini et al. ~\shortcite{franceschini03}, Elbaz et al. ~\shortcite{elbaz04}) and the link between the infrared and sub-mm populations at z$>$2 ~\cite{blain04} providing a definitive record of the cosmic star formation history and cosmic fossil background of the Universe.

\section{Acknowledgements}

The author wishes to thank the referee for comments and suggestions that improved the clarity of this work. Chris Pearson is supported by a European Commission International Co-operation Fellowship to Japan.



\bsp 

\label{lastpage}

\end{document}